
  \magnification\magstep 1

  \def\square{\kern1pt\vbox{\hrule height 0.5pt\hbox
  {\vrule width 0.5pt\hskip 2pt
  \vbox{\vskip 4pt}\hskip 2pt\vrule width 0.5pt}\hrule height
  0.5pt}\kern1pt}

  \font\bb=msbm10 scaled 1200

  \line{\hfil DAMTP/R-93/5}
  \line{\hfil CTP-TAMU-23/94}

  \vskip 0.4truecm
  \centerline{\bf Macroscopic superstrings as interpolating solitons}
  \vskip 1truecm
  \centerline{M.J. Duff${}^*$, G.W. Gibbons${}^{**}$ and P.K.
  Townsend${}^{**}$} \vskip 1 cm
  \centerline{${}^*$: Dept. of Physics, Texas A\& M Univ.,}
  \centerline{College Station, Texas, U.S.A.}
  \vskip 1cm
  \centerline{${}^{**}$:DAMTP, Univ. of Cambridge,}
  \centerline{Silver St., Cambridge, U.K.}

  \bigskip
  \bigskip
  \centerline{\bf ABSTRACT}
  \vskip 0.5cm
  It is known that, in string sigma-model metric, the `extreme' fivebrane
  solution of D=10 supergravity interpolates between D=10 Minkowski
  spacetime and a supersymmetric $S^3$ compactification to a linear
  dilaton vacuum. We show here that, in {\it fivebrane} sigma-model
  metric, the extreme string solution of D=10 supergravity
  interpolates between Minkowski spacetime and a hitherto unknown
  supersymmetric $S^7$ compactification of d=10 supergravity to a
  three-dimensional anti-de Sitter generalization of the linear dilaton
  vacuum, which may be invariantly characterized in terms of conformal
  Killing vectors. The dilaton field diverges near the string core but
  this divergence may be eliminated by re-interpreting the string
  solution as the extreme membrane solution of 11-dimensional
  supergravity. We show that the latter has an analytic extension
  through a regular degenerate event horizon to an interior region
  containing a curvature singularity. We obtain analogous results for
  other extended object solutions of supergravity theories.

  \vfill\eject

  \noindent{\bf 1. Vacuum interpolation via macroscopic superstrings}
  \vskip 0.5cm

  The action for the bosonic sector of $N=1$ $D=10$ supergravity,
  which can be viewed as (part of) the field theory limit of the
  heterotic string, can be written as
  $$
  S={1\over 2\kappa^2}\int d^{10}x\, \sqrt{-g} e^{-2\phi}[R +
  4(\partial\phi)^2 -{1\over12}F_3^2]
  \eqno (1.1)
  $$
  where $F_3$ is the three-form field-strength for a two-form
  potential, and $\phi$ is the dilaton field. There is an intrinsic
  ambiguity in the spacetime metric in theories with a dilaton because
  a new, although conformally equivalent, metric may be obtained by
  rescaling the old one by any positive function of the dilaton, e.g. by
  a power of $e^\phi$. A choice of metric in this conformal equivalence
  class corresponds to a choice of positive function of $\phi$
  multiplying the Einstein term in the action. The choice of
  $e^{-2\phi}$, as in (1.1), corresponds to the string sigma-model
  metric since this is the metric that couples to the worldsheet of the
  string. Non-singular spacetimes in this metric that solve the field
  equations may be regarded as approximate solutions of string theory.
  For the purposes of this paper we shall say that a spacetime with
  metric $g_{\mu\nu}$, dilaton $\phi$ and $q$-form field strength $F_q$
  is non-singular if
  \vskip 0.5cm
  (i) $g_{\mu\nu}$ is geodesically complete
  \vskip 0.5cm
  (ii) $\phi$ is everywhere finite and smooth
  \vskip 0.5cm
  (iii) $F_q$ has everywhere finite and smooth components in a
  coordinate system in which $g_{\mu\nu}$ has finite and smooth
  components.
  \vskip 0.5cm
  An example is the extreme `elementary' fivebrane solution [1]
  $$
  \eqalign{
  ds^2 &= -dt^2 + d{\bf x}\cdot d{\bf x} +
  \big[1-\left({a\over r}\right)^2\big]^{-2} dr^2 + r^2 d\Omega_3^2\cr
  e^{-2\phi} &= \big[1-\left({a\over r}\right)^2\big]\cr
  F_3 &=2a^2\varepsilon_3\ ,\cr}
  \eqno (1.2)
  $$
  where $d{\bf x}\cdot d{\bf x}$ is the Euclidean 5-metric,
  $d\Omega_3^2$ the standard metric on the unit 3-sphere and
  $\varepsilon_3$ its volume 3-form, and $a$ is a constant.  Evidence for
  the interpretation of this solution as an `extended soliton' is
  provided by a recent demonstration [2] that it interpolates between two
  supersymmetric vacua of $N=1$ $d=10$ supergravity, one being d=10
  Minkowski space and the other a previously unknown $S^3$
  compactification of $d=10$ supergravity to seven-dimensional Minkowski
  spacetime with a linear dilaton.

  A class of string-like solutions of the field equations of the action
  (1.1) was found in [3]. In the `extreme' limit of interest to us here,
  and expressed in terms of the seven form $F_7\equiv
  e^{-2\phi}\star F_3$, where $\star$ indicates the Hodge dual, the
  solution takes the form
  $$
  \eqalign{ ds^2 &= \big[1-\left( a\over
  r\right)^6\big][ -dt^2 + d\sigma^2] + \big[1-\left( a\over
  r\right)^6\big]^{-{5\over3}} dr^2 + r^2 \big[1-\left( a\over
  r\right)^6\big]^{1\over3} d\Omega_7^2\cr
  e^{-2\phi}&= \big[1-\left(a\over r\right)^6\big]^{-1}\cr
  F_7&= 6a^6\varepsilon_7 \cr}
  \eqno (1.3)
  $$
  where $d\Omega_7^2$ is the standard metric on the unit 7-sphere and
  $\varepsilon_7$ its volume 7-form. Unlike the spacetime metric of the
  fivebrane solution, the metric (1.2) is singular as $r\rightarrow a$.
  But this is the string sigma-model metric; the fact that a string in
  $D=10$ is dual to a fivebrane suggests that we instead consider
  the metric coupling to the six-dimensional worldvolume of a
  fivebrane, i.e. the {\it fivebrane sigma-model metric}, which is the
  conformally rescaled metric [4]
  $$
  d\tilde s^2 \equiv e^{-{2\over3}\phi}
  ds^2  = \big[1-\left( a\over r\right)^6\big]^{2\over3}[-dt^2
  +d\sigma^2] + \big[1-\left( a\over r\right)^6\big]^{-2} dr^2 + r^2
  d\Omega_7^2 \ , \eqno (1.4)
  $$
  This rescaled metric is still singular on the horizon at $r=a$ but
  now the singularity is merely due to an inappropriate
  choice of coordinates. As a first step towards this result, proved in
  section 3, we now investigate the asymptotic form of the metric near
  $r=a$. As in [2] we define
  $$
  \lambda = {6(r-a)\over a}
  \eqno (1.5)
  $$
  to get
  $$
  d\tilde s^2 = \big\{ \lambda^{2\over3}
  [-dt^2 + d\sigma^2] + {a^2\over 36}\lambda^{-2} d\lambda^2 + a^2
  d\Omega_7^2\big\}\big(1 + O(\lambda)\big)
  \eqno (1.6)
  $$
  Neglecting the $O(\lambda)$ terms, and defining
  $$
  \rho = {a\over6} {\rm ln} \lambda \ ,
  \eqno (1.7)
  $$
  we find that
  $$
  d\tilde s^2 \sim e^{4\rho\over a} (-dt^2 + d\sigma^2) +
  d\rho^2  + a^2 d\Omega_7^2
  \eqno (1.8)
  $$
  which is the standard metric on the product of $S^7$ with
  three-dimensional anti-de Sitter space, $adS_3$.

  The asymptotic form of the dilaton (near $r=a$) in the
  new coordinates is
  $$
  \phi \sim 3{\rho\over a}
  \eqno (1.9)
  $$ i.e. a
  {\it linear} dilaton. Clearly, the dilaton is linear only for the
  special choice of coordinates used here and since the spacetime factor
  of the asymptotic metric is not Minkowski (as it was for the fivebrane
  solution) but rather $adS_3$, the geometrical significance of
  linearity is unclear from the above result. We shall return to this
  point below, but it already clear from (1.9) that the dilaton
  diverges on the horizon at $r=a$, so that the extreme string
  {\it solution} of (1.3), as against just the {\it metric}, is singular
  there.

  If the action (1.1) is now rewritten in terms of the new, fivebrane
  sigma-model, metric, and with $F_7$ in place of $F_3$, it becomes
  $$
  S\rightarrow \tilde S = {1\over 2\kappa^2}\int d^{10}x\, \sqrt{-g}
  e^{{2\over3}\phi}\left[ R -{1\over 2.7!} F_7^2\right]
  \eqno (1.10)
  $$
  where $F_7$ may now be interpreted as the field strength of a
  six-form potential satisfying the Bianchi-identity $dF_7\equiv 0$. We
  have just shown that the extreme string solution of the field
  equations of this action interpolates between $D=10$ Minkowski
  spacetime and $adS_3\times S^7$ near the core, with a `linear' dilaton
  that diverges at $r=a$. In this respect the string solution in
  fivebrane sigma-model metric is similar to the fivebrane solution in
  string sigma-model metric, as string/fivebrane duality would suggest.
  We verify below that the $S^7$ compactification of D=10 supergravity
  implied by this analysis indeed exists, incidentally elucidating the
  geometrical significance of `linear' in the $adS$ context.

  In other respects, however, the string solution in fivebrane
  sigma-model metric is quite different from the fivebrane solution in
  string sigma-model metric. In particular, the string solution has an
  event horizon at which the dilaton diverges, so it fails to be a
  non-singular solution of $D=10$ supergravity, in the sense of this
  paper\footnote{${}^*$}{Whether this is a `physical' singularity is a
  more subtle question. In [4] it was shown using a test probe/source
  approach that if the test probe and source are both strings or both
  fivebranes, the probe falls into r=a in a finite proper time.  If one is a
  string and the other a fivebrane, however, it takes an infinite proper
  time.  So by these different criteria, the singularity structure is
  symmetrical between strings and fivebranes.}. However, we shall
  show that the  divergence of
  the dilaton at the horizon may be eliminated by re-interpreting the
  extreme string solution as the extreme membrane solution [5] of $D=11$
  supergravity. The latter solution still has an event horizon but now
  one can continue the solution analytically through it to a region
  containing a singularity at $r=0$. So even in $D=11$ the solution is
  not completely non-singular but is rather more analogous to the
  four-dimensional extreme Reissner-Nordstrom solution in which a
  singularity is hidden behind a degenerate event horizon.

  \vskip 1cm

  \noindent
  {\bf 2. The $S^7$ compactification}
  \vskip 0.5 cm

  Let $\{ x^M; \, M=0,1,\dots ,9\}$ be coordinates for the
  ten-dimensional spacetime, and define
  $$
  \Phi= e^{{2\over3}\phi}\ .
  \eqno (2.1)
  $$
  The field equations of (1.10) can now be written as
  $$
  R_{MN} = \Phi^{-1}D_MD_N\Phi
  + {1\over 2. 6!}F_{MP_1\dots P_6}F_N{}^{P_1\dots P_6} -
  {1\over 3.7!}g_{MN} F_7^2 \eqno (2.2a)
  $$
  $$
  \partial_M (\sqrt{-g}\Phi F^{MN_1\dots N_6})=0
  \eqno (2.2b)
  $$
  $$
  \square\Phi =  {1\over 3.7!}F_7^2 \Phi
  \eqno (2.2c)
  $$

  We now split the coordinates $\{x^M\}$ into $\{x^\mu, y^m\}$ with
  $\mu=0,1,2$ and $m=1,\dots, 7$, and seek product metrics of the form
  $$
  g_{\mu\nu}= g_{\mu\nu}(x) \qquad g_{mn} =g_{mn}(y)
  \eqno (2.3)
  $$
  with $g_{\mu n}=0$. In this case, $\sqrt{-g}=e_3(x) e_7(y)$ where
  $e_3$ and $e_7$ are the scalar density volume factors for the three
  and seven-dimensional spaces, respectively. It follows that (2.2b) is
  now solved by setting
  $$
  F^{m_1\dots m_7} = 3m(e_7)^{-1}\varepsilon^{m_1\dots m_7}
  \eqno (2.4)
  $$
  for some constant $m$, with all other components of $F_7$ vanishing,
  whereupon (2.2c) becomes
  $$
  (\square -3m^2)\Phi=0
  \eqno (2.5)
  $$
  We now suppose that $\Phi=\Phi(x)$, and further that
  $$
  D_\mu \partial_\nu \Phi ={1\over3} g_{\mu\nu} \square \Phi
  \eqno (2.6)
  $$
  or, equivalently, in view of (2.5),
  $$
  D_\mu \partial_\nu \Phi = m^2 g_{\mu\nu}\Phi\
  \eqno (2.7)
  $$
  which implies (2.5) and therefore supercedes it. Given (2.7), (2.2a)
  reduces to the two equations
  $$
  R_{\mu\nu} = -2m^2 g_{\mu\nu} \qquad R_{mn} = {3\over2} m^2 g_{mn}
  \eqno (2.8)
  $$
  with $R_{m\nu}=0$. These equations are solved by the standard
  invariant metrics on $adS_3$ and $S^7$ respectively. It remains to
  solve (2.7). We can choose coordinates $(t,\sigma, \rho)$ on $adS_3$
  such that the $adS_3$ metric is
  $$
  ds^2 =  e^{2m\rho}\left(-dt^2 + d\sigma^2\right) + d\rho^2
  \eqno (2.9)
  $$
  In these coordinates it is straightforward to verify that
  $$
  \phi = {3m\over 2}\rho
  \eqno (2.10)
  $$
  solves (2.7). We have therefore found an $S^7$ compactification of
  $D=10$ supergravity to $adS_3$ with a linear dilaton. Setting $m=2/a$
  we recover (1.8) and (1.9), found previously as an asymptotic limit of
  the extreme string solution.

  Observe now that (2.6) implies that $k= g^{\mu\nu}\partial\mu
  \Phi\,\partial_\nu $ is a {\it conformal Killing vector} of $adS_3$. We
  have shown above that an eigenfunction of the Dalembertian on $adS_3$
  with eigenvalue $3m^2$ is the potential for a conformal Killing vector
  of $adS_3$. A similar observation was made previously in the context of
  an $S^3$ compactification of $D=10$ supergravity [6] to $adS_7$.
  In fact, the $S^7$ compactification exhibited above is obtained from the
  solution found in [6] by the
  analytic continuation $m\rightarrow im$. We now see that
  the linear dilaton can be characterized in a coordinate-free way as
  proportional to the logarithm of a conformal Killing potential.

  \vskip 1 cm
  \noindent{\bf 3. Eleven-dimensional interpretation}
  \vskip 0.5cm

  The extreme string solution of $D=10$ supergravity theory has a
  (fivebrane sigma-model) {\it metric} which is non-singular (in
  appropriate coordinates) on the horizon $r=a$. As for the extreme
  Reissner-Nordstrom solution of $D=4$ Einstein-Maxwell theory, this
  metric can be continued through the horizon to $r<a$, but it would
  seem, on the face of it, that the dilaton field cannot be extended
  through $r=a$ because (1.3) tells us that $\phi\rightarrow \infty$ as
  the horizon is approached and that it is complex for
  $r<a$\footnote{$^*$}{This problem does not arise for the fivebrane
  solution in string conformal gauge because the metric in that case has
  no horizon.}. However, the field $\Phi$,
  defined in (2.1), is given by
  $$
  \Phi =\left[ 1-\left({a\over r}\right)\right]^{1\over3} \ ,
  \eqno (3.1)
  $$
  which certainly remains real and bounded as we pass through the
  horizon. Moreover, the dilaton $\phi$ appears in the action (1.10)
  through $\Phi$ {\it only}. It is known that 10-dimensional
  supergravity can be viewed as the dimensional reduction of
  11-dimensional supergravity and this allows a re-interpretation of
  $\Phi^2$ as a component of an 11-metric. We shall show that the
  above macroscopic string solution is in fact non-singular at $r=a$
  when interpreted as a solution of 11-dimensional supergravity.

  The 11-dimensional solutions of interest are
  $$
  \eqalign{
  ds_{11}^2 &= \Phi^2 (dx^{11})^2 + d\tilde s^2\cr
  \star F_4 &= 6a^6\varepsilon_7}
  \eqno (3.2)
  $$
  where $ds_{11}^2$ and $F_4$ are the metric and 4-form of
  11-dimensional supergravity (and $\star$ is the Hodge dual in 11
  dimensions) and $d\tilde s^2$ is the D=10 fivebrane sigma-model
  metric. In eleven dimensions there is no dilaton field and no
  possibility of conformal rescaling to avoid singularities, so the
  question of whether the eleven-dimensional fields are singular reduces
  to whether the metric of (3.2) is singular.

  We begin by examining the $S^7$ compactification discussed in
  section (2) from this new point of view. In this case $\Phi=
  e^{m\rho}$ so
  $$
  ds_{11}^2 = e^{2m\rho}(-dt^2 + d\sigma^2 + dx_{11}^2) + d\rho^2 +
  4m^{-2} d\Omega_7^2\ .
  \eqno (3.3)
  $$
  Since the first four terms in (3.3) give the metric of $adS_4$, it is
  clear that, at least locally, (3.3) is just the standard $adS_4\times
  S^7$ vacuum of eleven-dimensional supergravity. The ten-dimensional
  solution corresponding to (2.9) may be identified as the
  ten-dimensional hypersurface $x^{11}=0$. Since this hypersurface is
  invariant under the involution $x^{11}\rightarrow - x^{11}$, it has
  vanishing second fundamental form, i.e. it is a totally geodesic
  submanifold of the eleven-dimensional spacetime.

  The global structure of the spacetime (3.3) is governed by the
  behaviour of its $adS_4$ factor. The `horospherical'
  coordinates $\{\rho, t,\sigma, x^{11}\}$ do not cover all of $adS_4$
  but, rather, just half of it; in fact, they cover $adS_4/J$ where $J$
  is the antipodal map. To see why this is so, it is convenient to regard
  $adS_4$ as the quadric $Q$ in ${\bb R}^{3,2}$ given by
  $$
  (X^0)^2 + (X^4)^2 - (X^1)^2 - (X^2)^2 - (X^3)^2 = {1\over m^2}
  \eqno (3.4)
  $$
  One has
  $$
  \eqalign{
  m(X^4 -X^3) &= \Phi\ , \cr
  mX^0 &=  t\, \Phi\ , \cr
  mX^1 &= \sigma\, \Phi\ , \cr
  m(X^4 + X^3) &= \Phi^{-1} + (\sigma^2 +(x^{11})^2 -t^2)\Phi\ , }
  \eqno (3.5)
  $$
  The hypersurfaces $\rho={\rm const.}$ in $adS_4$ correspond to the
  intersection of the null hyperplanes, $N$, in ${\bb R}^{3,2}$,
  defined by
  $$
  m(X^4 - X^3) = e^{m\rho} = e^{{2\over3}\phi} = \Phi \ ,
  \eqno (3.6)
  $$
  with the quadric $Q$. However, as long as $\rho$ is taken to be real
  (corresponding by (2.10) to real $\phi$) there is no way to cover
  that half of $adS_4$ for which $X^4-X^3$ is negative. The boundary of
  the region for which $X^4-X^3$ is positive is a null hypersurface
  ${\cal H}$ in $adS_4$ obtained by intersecting the the quadric $Q$
  with the null hyperplane, $N_0$, passing through the origin,
  i.e. $X^4 - X^3 =0$. The Killing vectors ${\partial \over \partial t}$,
  ${\partial \over \partial \sigma}$ and ${\partial \over \partial
  x^{11}}$ are, respectively, timelike, spacelike and spacelike
  everywhere on $adS_4$ except on the null hypersurface ${\cal H}=
  Q\cap N_0$. On ${\cal H}$, which is a degenerate Killing horizon, and
  corresponds to the event horizon $r=a$ in the extreme string solution,
  these Killing vectors are null. Note that the null hyperplane $N_0$
  divides $adS_3$ into two isometric pieces, that for which $X^4>X^3
  \Leftrightarrow \Phi>0$ and that for which $X^4<X^3 \Leftrightarrow
  \Phi<0$. These two halves are interchanged by the antipodal
  involution $J: X^A \rightarrow - X^A\, A=0,1,2,3,4$. Evidently, from
  the point of view of the geometry of $adS_4$ there is no reason for
  $\phi$ to be real or finite, nor for $\Phi$ to be positive.

  Note that on $adS_4$ the horizon ${\cal H}$ has two connected
  components, ${\cal H}^+$ and ${\cal H}^-$, corresponding to the two
  values of $X^0$ satisfying (3.4) with $\Phi=m(X^4-X^3)=0$. One
  component, ${\cal H}^-$, is the past horizon and one component,
  ${\cal H}^+$, is the future horizon. On the universal covering space
  $\widetilde{adS}_4$ there are infinitely many such pairs. On the
  identified space $adS_4/J$ the two components ${\cal H}^+$ and
  ${\cal H}^-$ are identified.

  The above discussion for $adS_4$ may readily be extended to the
  string metric (1.4). In this case $\Phi$ is given by (3.1) and the
  corresponding eleven-dimensional metric is
  $$
  ds_{11}^2 = \left[1-\left(a\over r\right)^6\right]^{2\over3}\left[
  -dt^2 + d\sigma^2 + (dx^{11})^2 \right] +
  \left[1-\left(a\over r\right)^6\right]^{-2}dr^2 + r^2 d\Omega_7^2
  \eqno (3.7)
  $$
  Observe that this metric can be interpreted as that of a {\it
  membrane} rather than a string. In fact, it is just the membrane
  solution of [5] but written in terms of our Schwarzschild-type radial
  coordinate $r$ rather than the isotropic radial coordinate of [5],
  which we shall call $\hat r$; the relation between the two is $\hat r
  = (r^6 - a^6)^{1\over6}$. In [5] this solution was interpreted as the
  metric exterior to a singular membrane source located at $\hat r=0$,
  i.e. at the horizon. Fortunately one need not
  suppose the existence of such a source since the metric can be smoothly
  extended through $r=a$ by choice of a suitable set of coordinates,
  although, like the extreme Reissner-Nordstrom black hole, there is a
  singularity hidden behind the horizon, at $r=0$. One way of showing
  this is to follow closely the discussion for the $adS_4$ case. There,
  the four coordinates $\{X^0,X^1, X^2, X^3\}$ (defined in terms of
  $\{\Phi, t,\sigma, x^{11}\}$ by eqs. (3.5)), cover the horizon ${\cal
  H}$, with $\Phi= e^{m\rho}$; the field $\Phi$ is a well-behaved
  function on $adS_4$ and one may check that the metric written in terms
  of the coordinates $\{X^0, X^1, X^2, X^3\}$ is well-behaved in a
  neighbourhood of the horizon. In the case of the metric (3.7) one can
  still use  $\{X^0,X^1, X^2, X^3\}$ defined in terms of
  $\{\Phi,t,\sigma, x^{11}\}$ by (3.5) but now one uses (3.1) to express
  this metric in terms of $\{X^0,X^1, X^2, X^3\}$. Thus we set
  $$
  r= a (1-\Phi^3)^{-{1\over 6}}
  \eqno (3.8)
  $$
  and re-express (3.7) in
  terms of $(\Phi,t,\sigma, x^{11})$ to get
  $$
  \eqalign{
  ds_{11}^2 = \bigg\{\Phi^2[-dt^2 &+
  d\sigma^2 + (dx^{11})^2] + 4a^2 \Phi^{-2} d\Phi^2
  + a^2 d\Omega_7^2 \bigg\}\cr
  & + a^2\big[(1-\Phi^3)^{-{1\over3}}-1\big]
  \big[4\Phi^{-2}d\Phi^2 + d\Omega_7^2\big] \ .}
  \eqno (3.9)
  $$
  That is,
  $$
  ds^2_{11} = ds^2_{11}({\rm asymptotic}) +
  a^2\big[(1-\Phi^3)^{-{1\over3}}-1\big] \big[4\Phi^{-2}d\Phi^2 +
  d\Omega_7^2\big]
  \eqno (3.10)
  $$
  where the first term on the right hand side is asymptotic metric as
  $r\rightarrow a$, i.e. the standard metric on $adS_4\times S^7$. Since
  the function $\Phi=m(X^4-X^3)$ is an analytic function on $adS_4$ the
  remaining terms in (3.10) extend analytically through the horizon
  ${\cal H}$, at $r=a$, to give an analytic metric up to the
  curvature singularity at $\Phi=-\infty$, i.e. $r=0$.

  The local chart we have introduced contains two connected components
  of the horizon, ${\cal H}^+$ and ${\cal H}^-$. If one continues the
  extension to obtain the maximal analytic simply-connected extension,
  there will be, just as for $\widetilde{adS}_4$, infinitely many pairs
  of such horizons. The Carter-Penrose diagram obtained by supressing the
  angular coordinates on $S^7$ and the two ignorable worldsheet
  coordinates $\sigma$ and $x^{11}$ resembles that of the extreme
  Reissner-Nordstrom metric.

  It follows from the above analysis, by simply setting
  $x^{11}=0$, that the $D=10$ metric of (1.4) can also be extended
  analytically through the horizon to an interior region with
  a singularity at $r=0$, so we have now established that the
  singularity of this metric at $r=a$ is merely a coordinate
  singularity. The higher-dimensional interpretation is required
  only to circumvent the singularity of the $D=10$ dilaton
  field on the horizon

  \vskip 1cm
  \noindent{\bf 4. Generalization to super $p$-branes}
  \vskip 0.5cm
  We have seen that the ten-dimensional macroscopic superstring
  interpolates between the flat ten-dimensional Minkowski vacuum and a
  compactification on $S^7$ to $adS_3$ with a `linear' dilaton, which
  is invariantly characterized in terms of a conformal Killing
  potential. We also saw that the problem of the divergence of the
  dilaton at the horizon could be circumvented by passing to a
  non-singular 11-dimensional spacetime, re-interpreting the solution
  as a membrane. We shall now investigate to what extent these
  features persist in a class of metrics representing macroscopic
  $p$-branes in a $D$-dimensional spacetime. We consider an action of
  the form
  $$
  S={1\over 2\kappa^2}\int d^Dx\, \sqrt{-g}\left[ R
  -{1\over2} (\partial\phi)^2 -{1\over
  2(d+1)!}e^{-\alpha\phi}F_{d+1}^2\right]
  \eqno (4.1)
  $$
  where $F_{d+1}$ is a field-strength for a d-form potential $A$, and
  $\alpha$ is a constant.
  This action is expressed in terms of a metric
  that we shall call the Einstein metric, since the Einstein term is the
  canonical one, without factors of $e^\phi$. The action in terms of the
  $(d-1)$-brane sigma-model metric, i.e. the metric coupling to the
  d-dimensional worldvolume of such an object, is obtained by the
  conformal rescaling
  $$
  g_{\mu\nu} \rightarrow e^{-{\alpha\over d}\phi} g_{\mu\nu}\ ,
  \eqno (4.2)
  $$ which yields
  $$
  \eqalign{
  S={1\over 2\kappa^2}\int d^Dx\, \sqrt{-g} e^{-{(D-2)\alpha\over
  2d}\phi} \big[ R - &{1\over2}\left(1 -{\alpha^2(D-1)(D-2)\over
  2d^2}\right) (\partial\phi)^2 \cr
  &-{1\over 2(d+1)!}F_{d+1}^2\big]\ .\cr}
  \eqno (4.3)
  $$
  For $\alpha$ satisfying
  $$
  \alpha^2 = 4 - {2d\tilde d\over (D-2)}
  \eqno (4.4)
  $$
  extreme $(\tilde d -1)$-brane solutions of the equations of motion of
  (4.1) are given in [7], where
  $$
  \tilde d = D-d-2\ .
  \eqno (4.5)
  $$
  After the rescaling of (4.2) these
  solutions are
  $$
  \eqalign{
  ds^2 &= \Delta^{d-2\over d}\left(-dt^2 +d{\bf x}\cdot d{\bf x}\right)
  + \Delta^{-2} dr^2 + r^2 d\Omega_{d+1}^2\cr
  e^{-2\phi} &=\Delta^\alpha\cr
  F_{d+1} &= da^d\varepsilon_{d+1}\cr}
  \eqno (4.6)
  $$
  where $d{\bf x}\cdot d{\bf x}$ is the Euclidean $(\tilde d
  -1)$-metric, and
  $$
  \Delta = 1-\left(a\over r\right)^d\ .
  \eqno (4.7)
  $$
  For $D=10$ these solutions include the fivebrane and string solutions
  previously discussed.

  To determine the asymptotic form of the
  metric of (4.6) near $r=a$ we let
  $$
  r= a\left( 1+{\lambda\over d}\right) \ ,
  \eqno (4.8)
  $$
  in which case
  $$
  ds^2 = \left[ \lambda^{d-2\over d}(-dt^2 +d{\bf x}\cdot d{\bf x})
  +\left({a\over d}\right)^2\lambda^{-2}d\lambda^2 + a^2
  d\Omega_{d+1}^2\right]\left(1+O(\lambda)\right)\ .
  \eqno (4.9)
  $$
  Neglecting the $O(\lambda)$ terms, as before, and defining the
  new coordinate $\rho$ by
  $$
  \lambda = e^{{d\over a}\rho}
  \eqno (4.10)
  $$
  we get
  $$
  \eqalign{
  ds^2 &\sim e^{{(d-2)\over a}\rho}(-dt^2 +d{\bf x}\cdot d{\bf x}) +
  d\rho^2 + a^2d\Omega_{d+1}^2\cr
  \phi &\sim -{d\alpha\over 2a}\rho\cr
  F_{d+1} &\sim da^d \varepsilon_{d+1}\ .\cr}
  \eqno (4.11)
  $$
  This agrees with (1.8) and (1.9) for the extreme string solution of
  $D=10$ supergravity discussed in section 1, for which
  $\alpha=-1$\footnote{$^*$}{And corrects a factor of 2 error in the
  result of [2] for the extreme fivebrane solution, for which
  $\alpha=1$; note that $\alpha\rightarrow -\alpha$ under duality.}

  If $d=2$ (for which $\alpha \neq 0$), then (4.11) reduces to
  $$
  \eqalign{
  ds^2 &\sim (-dt^2 +d{\bf x}\cdot d{\bf x} +
  d\rho^2) + a^2d\Omega_{d+1}^2\cr
  \phi &\sim -{\alpha\over a}\rho\cr
  F_3 &\sim 2a^2 \varepsilon_{3}\cr}
  \eqno (4.12)
  $$
  which is $(Mink)_{\tilde d +1} \times S^3$, with a linear dilaton
  vacuum. That is, in string sigma-model metric the
  dual $(\tilde d -1)$-brane interpolates between $D$-dimensional
  Minkowski spacetime and the product of $S^3$ with a $(\tilde d
  +1)$-dimensional {\it Minkowski} spacetime, generalizing the case of
  the $D=10$ fivebrane, which was noted in [2] to have this property.
  For these cases the asymptotic behaviour of the dilaton near the
  p-brane core can be invariantly characterized as linear in an ignorable
  coordinate associated with a space-translation Killing vector. Since
  there is no event horizon the $(\tilde d-1)$-brane solution is
  completely non-singular.

  If $d\ne 2$ the asymptotic spacetime spacetime is $(adS)_{\tilde
  d+1}\times S^{d+1}$, and there is an event horizon at $r=a$. In terms
  of the new coordinate
  $$
  z= -{2a\over (d-2)} e^{-{(d-2)\over 2a}\rho}\ ,
  \eqno (4.13)
  $$
  the asymptotic adS metric near the horizon is
  $$
  \left( {2a\over (d-2)z}\right)^2 \left(-dt^2 +
  d{\bf x}\cdot d{\bf x} + dz^2\right)\ .
  \eqno (4.14)
  $$
  Clearly, the vector
  $$
  k= \left({d-2\over 2a}\right){\partial\over \partial z}=
  {\partial\over \partial \rho}\left( e^{{(d-2)\over 2a}\rho}\right)\,
  {\partial\over \partial \rho}
  \eqno (4.15)
  $$
  is a conformal Killing vector with conformal Killing potential
  $$
  \Phi = e^{{(d-2)\over 2a}\rho} = e^{-{(d-2)\over d\alpha}\phi}\ ;
  \eqno (4.16)
  $$
  that is
  $$
  k^\mu = g^{\mu\nu}\partial_\nu \Phi \ .
  \eqno (4.17)
  $$
  This provides an invariant characterization of the `linear
  dilaton' vacuum in the general $d\ne 2$ case.

  The solutions just described for $d\ne 2$ cannot be considered as
  non-singular in $D$-dimensions because the dilaton is singular on the
  event horizon. We now investigate whether this problem can be
  circumvented by re-interpreting the solution as a $\tilde d$-brane in a
  $(D+1)$-dimensional spacetime, as was possible for the
  $D=10$ macroscopic superstring. From (4.3) we see that this is
  certainly possible if
  $$
  \alpha^2 = {2d\over (D-1)(D-2)}
  \eqno (4.18)
  $$
  because the absence of a $(\partial\phi)^2$ term allows an
  immediate interpretation in $(D+1)$ dimensions with
  $\Phi^2$ as the component of the metric in the extra dimension. This
  is compatible with the restriction on $\alpha$ assumed
  for the class of solutions being considered here, eq. (4.4), when $D$
  satisfies
  $$
  D={(d-1)(d+2)\over (d-2)}= (d+3) +{4\over (d-2)}\ ,
  \eqno (4.19)
  $$
  which allows only $D=9$ and $D=10$. The $D=9$ case occurs for $d=4$,
  i.e. $\tilde d =5$, so this case represents a dilatonic $D=9$
  membrane which, as we have now shown, can be reinterpreted as a $D=10$
  threebrane, in fact as the selfdual threebrane of Horowitz and
  Strominger [8]. There are two $D=10$ cases. One occurs for $d=6$ and
  is just the macroscopic string solution already discussed. The other
  occurs for $d=3$, i.e. $\tilde d = 5$, so this case represents a
  dilatonic $D=10$ fourbrane which we see can be interpreted as a
  $D=11$ fivebrane. In fact, it is G{\" u}ven's fivebrane solution of
  $D=11$ supergravity [9].

  In each of these cases the $(D+1)$-metric
  takes the form
  $$
  \eqalign{
  ds^2 = &\big[1- \left({a\over r}\right)^{(D-p-2)}\big]^{1\over
  p+1}\big[ -dt^2 + d{\bf x}\cdot d{\bf x} + (dx_*)^2\big]\cr
  &+ \big[1- \left({a\over r}\right)^{(D-p-2)}\big]^{-2} + r^2
  d\Omega_{d+1}^2\ ,}
  \eqno (4.20)
  $$
  where $p$ is the dimension of the extended object (now $\tilde d$
  rather than $\tilde d -1$) and $x_*$ is the coordinate of the extra
  dimension of  spacetime. For $d>1$ the only other way to re-interpret
  the action in $(D+1)$ dimensions is by interpreting the $d$-form
  potential as a mixed component of a $d+1$ form potential in $(D+1)$
  dimensions and this has the effect of replacing $d$ by $\tilde d$ on
  the right hand side of (4.18) and (4.19), and hence allows the same
  values of $D$. The $(D+1)$-dimensional solutions are the same as
  above\footnote{$^*$}{If $d=1$ the potential $A$ is a one-form and
  there arises the possibility of a Kaluza-Klein interpretation of it as
  an off-diagonal component of the $(D+1)$-metric. This allows other
  values of $D$, e.g. $D=4$, but these cases are rather dissimilar to the
  $D=10$ macroscopic superstring and will not be discussed here.}.

  The metrics (4.20) belong to a general class of $p$-brane metrics
  for arbitrary $p$ and general spacetime dimension, although only a
  few are solutions of supergravity theories. All have an event
  horizon and using the techniques described earlier for the macroscopic
  string one can show that they can all be analytically
  continued through the horizon. In those cases for which $p$ is even
  the extension through the horizon is very similar to that described
  for the eleven-dimensional membrane ($p=2$) in section 3; in each
  case there is a singularity in the interior region at $r=0$. For $p$
  odd, however, the extension is quite different. As will be shown
  elsewhere [10], the interior region
  is isometric to the exterior one when $p$ is odd and so there is no
  singularity there! Thus, unlike the $D=11$ membrane, the $D=10$
  three-brane and $D=11$ fivebrane are {\it completely non-singular},
  despite the presence of an event horizon. Similar remarks apply to the
  self-dual string [7] in $D=6$.

  \vskip 2cm
  \centerline{\bf References}
  \vskip 1cm
  \item {[1]}
  M.J. Duff and X. Lu, Nucl. Phys. {\bf B354} (1991) 141;
  \item {[2]}
  G.W. Gibbons and P.K. Townsend, Phys. Rev. Lett. {\bf 71} (1993) 3754.
  \item {[3]}
  A. Dabholkar, G.W. Gibbons, J.A. Harvey and F. Ruiz-Ruiz,  Nucl. Phys.
  {\bf B340} (1990) 33.
  \item {[4]}
  M.J. Duff, R.R. Khuri and J.X. Lu, Nucl. Phys. {\bf B377} (1992) 281.
  \item {[5]}
  M.J. Duff and K.S. Stelle, Phys. Lett. {\bf B253} (1991) 113.
  \item {[6]}
  M.J. Duff, P.K. Townsend and P. van Nieuwenhuizen, Phys. Lett.
  {\bf 122B} (1983) 232.
  \item {[7]}
  M.J. Duff and X. Lu, Nucl. Phys. {\bf B411} (1994) 301.
  \item {[8]}
  G. Horowitz and A. Strominger, Nucl. Phys. {\bf B360} (1991) 197.
  \item {[9]}
  R. G{\" u}ven, Phys. Lett. {\bf B276} (1992) 49.
  \item {[10]}
  G.W. Gibbons, G. Horowitz and P.K. Townsend, in preparation.

  \end